\documentclass[preprint]{vgtc}               

\onlineid{1128}

\vgtccategory{Empirical Study}

\usepackage{xargs}      
\usepackage{soul}       
\usepackage{color}      
\usepackage{xspace}     
\usepackage{xpunctuate} 
\usepackage{booktabs}
\usepackage{amsmath, amsfonts}
\usepackage{multirow}
\usepackage{algorithmic}
\usepackage{algorithm}
\usepackage{array}
\usepackage[caption=false,font=normalsize,labelfont=sf,textfont=sf]{subfig}
\usepackage{textcomp}
\usepackage{stfloats}
\usepackage{url}
\usepackage{verbatim}
\usepackage{graphicx}
\usepackage{wrapfig}
\usepackage{mathptmx}                  
\usepackage{times}                     

\graphicspath{{figs/}{figures/}{pictures/}{images/}{./}} 

\newcommand{\ie}{{i.e.,}\xspace}
\newcommand{\eg}{{e.g.,}\xspace}
\newcommand{\etal}{{et~al\xperiod}\xspace}

\newcommand{\etc}{{etc\xperiod}\xspace}

\newcommand{\bpstart}[1]{\vspace{0px}\noindent\textbf{#1}}

\definecolor{lightpink}{RGB}{237,157,202}
\definecolor{lightred}{RGB}{210,121,121}
\definecolor{lightorange}{RGB}{230,170,50}
\definecolor{lightgold}{RGB}{210,194,121}
\definecolor{lightgreen}{RGB}{121,210,121}
\definecolor{lightaqua}{RGB}{121,206,210}
\definecolor{lightblue}{RGB}{121,124,210}
\definecolor{lightpurple}{RGB}{153,102,255}
\definecolor{red}{RGB}{178,34,34}
\definecolor{gray}{RGB}{166,166,166}

\newcommandx{\guest}[3][1=]
    {\setulcolor{lightorange}{\ul{#1}} \textcolor{lightorange} 
    {[\textbf{#2:} #3]}}
\newcommandx{\luke}[2][1=] 
    {\setulcolor{lightred}{\ul{#1}} \textcolor{lightred} 
    {[\textbf{Luke:} #2]}}  
\newcommandx{\chenglong}[2][1=] 
    {\setulcolor{lightpurple}{\ul{#1}} \textcolor{lightpurple}  
    {[\textbf{Chenglong:} #2]}}
\newcommandx{\steven}[2][1=] 
    {\setulcolor{lightblue}{\ul{#1}} \textcolor{lightblue}  
    {[\textbf{Steven:} #2]}}

\newcommand{\figureheatmap}{
    \begin{figure}[!t]
        \centering
        \resizebox{0.90\linewidth}{!}{\includegraphics{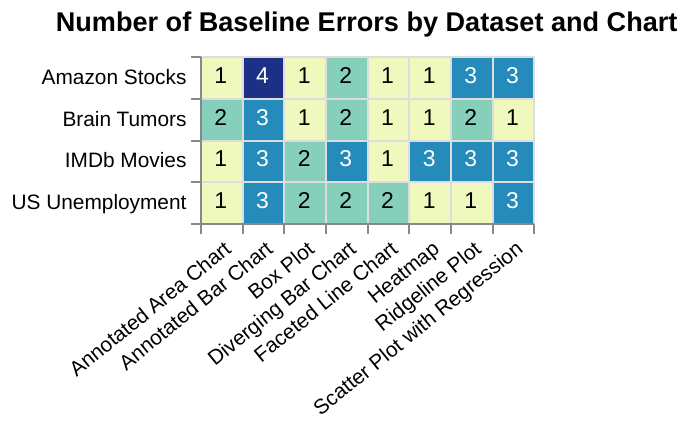}}
        \vspace{-15pt}
        \caption{
            Baseline retargeting errors across 32 different chart instances.
            Higher errors are observed for more complex charts (annotations, regression) and datasets with high (IMDb) or low (US Unemployment) complexity.
        }
        \label{fig:heatmap}
        \vspace{-5mm}
    \end{figure}
}

\newcommand{\figureexamples}{
    \begin{figure*}[!t]
        \centering
        \resizebox{0.90\linewidth}{!}{\includegraphics{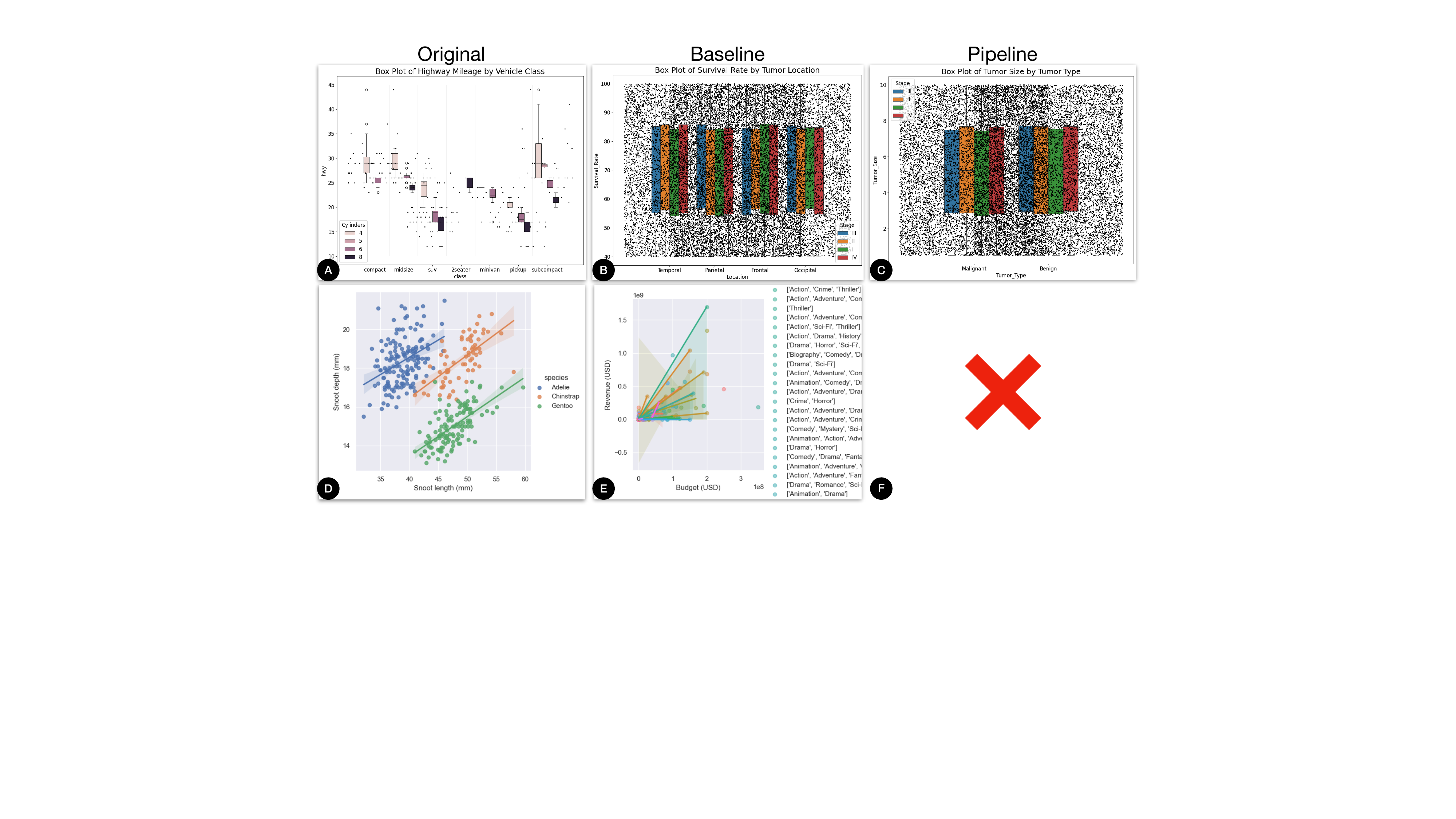}}
        \vspace{-8pt}
        \caption{
            Example retargeted chart specifications for both LLM baseline and pipeline. (A-C) Retargeted Matplotlib box plot to brain tumor dataset. (D-F) Retargeted Seaborn scatter plot with regression to IMDb Movies dataset; the LLM pipeline fails to render the chart (F).
        }
        \label{fig:examples}
        \vspace{-5mm}
    \end{figure*}
}

\newcommand{\tableerroranalysis}{
    \begin{table}[!t]
    \centering
    \setlength\tabcolsep{0pt} 
    \smallskip 
    \small
    
    \def\arraystretch{0.90}%
    \scalebox{0.90}{
    \begin{tabular*}{\linewidth}{@{\extracolsep{\fill}} lllll}
        Error Class & Type & Severity & Baseline (\%) & Pipeline (\%) \\
        
        \toprule 

        \multirow{3}{*}{Syntactic} & Data transformation not valid & High & 21.88 & 18.75 \\
        & Old data field(s) not removed & High & 0.0 & 28.13 \\
        & Library import removed & High & 0.0 & 3.13 \\

        \midrule
        \multirow{6}{*}{Semantic} & Incorrect / inappropriate scales & Medium & 16.0 & 12.5 \\
        & Incorrect axis label(s) & Medium & 4.0 & 6.25 \\
        & Non-standard data encoding & Medium & 20.0 & 25.0 \\
        & Annotation not removed & Medium & 16.0 & 18.75 \\
        & Missing necessary transformation & Medium & 32.0 & 31.25 \\
        & Incorrect or ambiguous axis title & Medium & 12.0 & 6.25 \\

        \midrule
        \multirow{9}{*}{Pragmatic} & Overplotting & High & 60.0 & 75.0 \\
        & Unnecessary use of faceting & Low & 4.0 & 0.0 \\
        & Axis title missing units & Medium & 20.0 & 25.0 \\
        & Heavy whitespace & Low & 4.0 & 0.0 \\
        & Repeated axis labels & Low & 4.0 & 6.25 \\
        & Axis labels not rotated & Medium & 0.0 & 12.5 \\
        & Data encoding not necessary & Medium & 4.0 & 0.0 \\
        & Missing axis titles & Medium & 4.0 & 0.0 \\
        & Missing legend title & Medium & 8.0 & 6.25 \\

        \bottomrule
    \end{tabular*}
    }
    \vspace{4pt}
    \caption{ 
        Retargeting errors by type and prevalence (\%) for both LLM baseline and pipeline approaches, normalized by rendered charts.
    }
    \label{table:errors}
    \vspace{-6mm}
    \end{table}
}

\vgtcinsertpkg

\ieeedoi{10.1109/VIS60296.2025.00034}

\title{Challenges \& Opportunities with LLM-Assisted Visualization Retargeting}

\author{
    Luke S. Snyder\thanks{e-mail: snyderl@cs.washington.edu}\\ %
        \scriptsize University of Washington %
    \and Chenglong Wang\thanks{e-mail: chenglong.wang@microsoft.com}\\ %
        \scriptsize Microsoft Research %
    \and Steven M. Drucker\thanks{e-mail: sdrucker@microsoft.com}\\ %
        \scriptsize Microsoft Research
}

\teaser{
    \centering
    \includegraphics[width=0.95\linewidth]{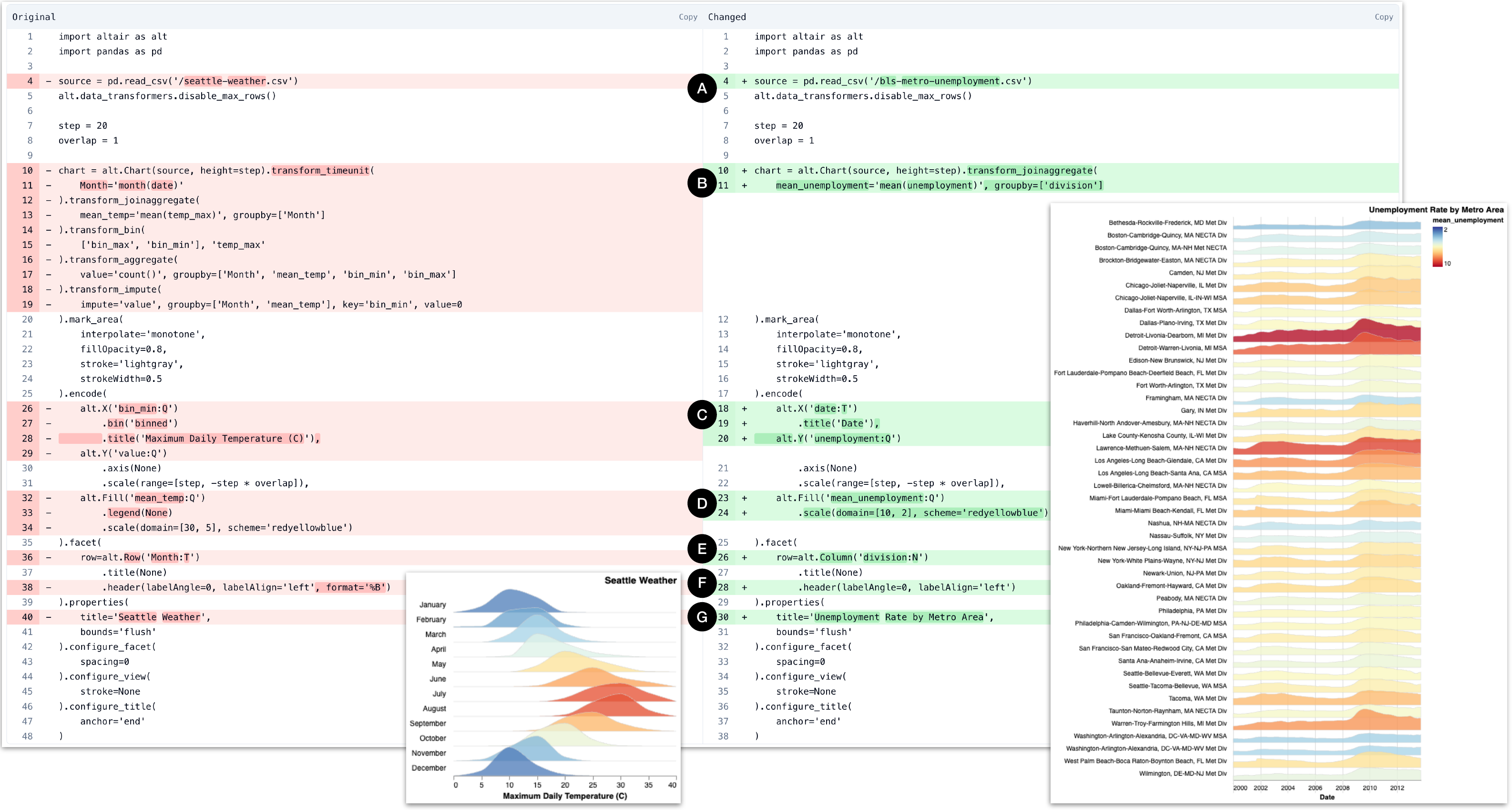}
    \caption{ 
        Manually retargeting an Altair ridgeline plot of Seattle weather data (left) to a new dataset containing United States unemployment rates across different metro areas (right). Summary of retargeting actions: (A) Providing the new dataset source; (B) Removing stale data transformations and replacing old data fields with new fields; (C-E) Updating new fields, domains, and titles for encoding channels; (F) Removing a stale formatter used for the month field; (G) Updating the chart title.
    }
    \label{fig:teaser}
}

\abstract{
    Despite the ubiquity of visualization examples published on the web, retargeting existing custom chart implementations to new datasets remains difficult, time-intensive, and tedious.
The adaptation process assumes author familiarity with both the implementation of the example as well as how the new dataset might need to be transformed to fit into the example code. 
With recent advances in Large Language Models (LLMs), automatic adaptation of code can be achieved from high-level user prompts, reducing the barrier for visualization retargeting.
To better understand how LLMs can assist retargeting and its potential limitations, we characterize and evaluate the performance of LLM assistance across multiple datasets and charts of varying complexity, categorizing failures according to type and severity.
In our evaluation, we compare two approaches: (1) directly instructing the LLM model to fully generate and adapt code by treating code as text inputs and (2) a more constrained program synthesis pipeline where the LLM guides the code construction process by providing structural information (\eg visual encodings) based on properties of the example code and data.
We find that both approaches struggle when new data has not been appropriately transformed, and discuss important design recommendations for future retargeting systems.

} 


\begin{document}



\maketitle

\section{Introduction}

Data visualizations are typically authored from scratch or adapted from extant designs.
A variety of tools support the former, ranging from low-level APIs~\cite{bostock2011d3} and interfaces for manual design~\cite{ren2018charticulator} to higher-level grammars~\cite{satyanarayan2016vega, satyanarayan2015reactive, heer2023mosaic} and toolkits~\cite{drucker2015unifying, stolte2008polaris} for rapid construction.
However, depending on the visualization task, analysis and communication goals, and designer expertise, authoring a visualization from scratch can be tedious, time-consuming, and difficult.
Consequently, charting tools often showcase design galleries ~\cite{bostock2011d3, satyanarayan2016vega, satyanarayan2015reactive} which may serve as useful starting points from which the designer can adapt and refine based on their own data and goals~\cite{bako2022understanding}.

We define \textit{Visualization Retargeting} as the process of adapting an existing chart to a new dataset.
Retargeting in the visualization community has typically focused on \textit{restyling} image outputs of charts, such as Scalable Vector Graphics (SVG)~\cite{cui2021mixed, harper2014deconstructing, harper2017converting} or bitmaps~\cite{poco2017reverse}.
These methods rely on deconstruction heuristics that are limited by what can feasibly be inferred from image output alone~\cite{snyder2023divi, chen2023mystique, poco2017reverse, masson2023chartdetective}, and lack context regarding how the source data was manipulated or transformed.
Recent advances in Large Language Models (LLMs) open the possibility of improving this restyling effort on code (rather than image output) by being able to reason about the semantics and context of the new data and how it might be appropriately fit to an existing visual design.
Critically, we explore the broader capabilities of LLM-assisted retargeting with code implementations of charts, encompassing not only possible restyling, but also transformations and mappings of new data (\eg filtering, aggregation), reuse and generation of visual scales and channels (\eg scale domains, legends), and modification of axis and chart guides (\eg titles, labels).

The primary contributions of this paper are threefold: (1) we outline and characterize the challenges that arise when retargeting an implemented visualization specification to new data; (2) we evaluate the performance of LLM-assisted retargeting across a variety of datasets and charts; and (3) we summarize opportunities and design recommendations for future retargeting systems to mitigate existing pitfalls with LLM assistance.
Our evaluation tests both an LLM baseline and LLM program synthesis pipeline that guides the code generation and refinement process by extracting salient metadata (\eg visual encodings and transforms), providing additional context to aid reasoning.
We find that the pipeline performs worse than the baseline, possibly due to propagating errors, and that both approaches are unable to appropriately retarget most charts unless the data has been adequately transformed and filtered in advance.

\section{Related Work}
\subsection{Chart Deconstruction \& Restyling}
Prior work in chart deconstruction primarily aims to reverse-engineer data from an SVG or bitmap visualization, which may be used to support interactive features~\cite{snyder2023divi}, restyling and redesign~\cite{harper2014deconstructing, harper2017converting}, or layout reuse~\cite{chen2023mystique, cui2021mixed}.
Deconstruction algorithms typically rely on heuristics that are limited in scope, such as rectangle charts~\cite{chen2023mystique} or infographics~\cite{cui2021mixed}, to ease inference.
Computer vision algorithms may also be used for bitmap images~\cite{poco2017reverse}, but are error-prone in certain cases (\eg overlapping marks~\cite{masson2023chartdetective}) without the additional structure and context that SVG format provides.

Our work draws on chart deconstruction methods by attempting to guide LLMs in detecting data mappings, scales, and encodings included in a chart specification.
This information provides necessary scaffolding to then map to a new dataset.
Given the wide range of possible errors with reverse-engineering a chart image, we instead focus on the chart's \textit{specification} format (\eg Vega-Lite~\cite{satyanarayan2016vega} grammar representation or Matplotlib Python code).
Specifications also provide important context that would be impossible in many cases to infer from an image alone, such as complex data transformations (\eg aggregation, filtering, \etc).

\subsection{Visualization Examples \& Automated Support}
Charts may be automatically constructed with smart defaults that depend on data types and semantics (\eg Show Me~\cite{mackinlay2007show}, Vega-Lite~\cite{satyanarayan2016vega}), although additional effort is required for customization.
Visualization examples and online templates are useful starting points for designers to adapt more complex designs to their own data, but take considerable time and care and require adequate familiarity with the tool~\cite{bako2022understanding}.
To aid this process, Shen \etal~\cite{shen2022galvis} proposed an automated pipeline to decouple Vega-Lite examples from their data source, which could then be modified with data fields from a new dataset.
We apply this line of work to LLMs, which enables broader coverage of visualizations across different tools and can support more complex retargeting operations (\eg data transformation, code generation, \etc).

\section{Challenges with Visualization Retargeting}
\label{sec:challenges}

Retargeting visualization code to a new dataset involves multiple decisions about how to both modify the chart specification itself and refine the visual design based on data semantics.
Using Fig.~\ref{fig:teaser} as an example, we identify and characterize different tiers of retargeting challenges across Kosslyn's~\cite{kosslyn1989understanding} categories for chart design and comprehension: syntax, semantics, and pragmatics.

\vspace{0.65mm}
\bpstart{Syntax.}
Syntax challenges when modifying code arise due to dependencies (\eg variables, data fields) within the specification.
For example, when retargeting the Altair ridgeline plot (Fig.~\ref{fig:teaser}), the designer must replace the dataset location (Fig.~\ref{fig:teaser}(A)) and data field names (Fig.~\ref{fig:teaser}(B-E)).
While these modifications are simple substitutions, more complex syntax changes are required when removing or updating data transformations or encodings.
For instance, when the designer chooses to encode \texttt{date} on the x-axis instead of \texttt{bin\_min} (Fig.~\ref{fig:teaser}(C)), transformations that derive \texttt{bin\_min} (Fig.~\ref{fig:teaser}(B)) are rendered stale and should be removed.
However, removing these transformations affects the derived field \texttt{value} that is used to encode the y-axis, which must then be replaced with another available field from the new dataset (Fig.~\ref{fig:teaser}(C)).
Additionally, the decision to facet by \texttt{division} in the new dataset (Fig.~\ref{fig:teaser}(E)) also makes the \texttt{transform\_timeunit} derivation obsolete (Fig.~\ref{fig:teaser}(A)).

Often, updating an encoding channel with a new data field also necessitates checking the formatting syntax, which may depend on the data type.
For example, replacing \texttt{Month} with \texttt{division} (Fig.~\ref{fig:teaser}(E)) requires the author to update the type information as well: \texttt{Month:T} (for temporal) to \texttt{division:N} (for nominal).
In addition, the formatting option for \texttt{month} (\texttt{format=`\%B'}) must be removed (Fig.~\ref{fig:teaser}(F)) for the visualization to render correctly with the categorical type of \texttt{division}.
Given that these dependencies vary between tools, LLMs can ostensibly help generate code that not only accounts for the typing and semantics of the new data, but is also syntactically correct for the charting tool at hand (\ie gulf of execution~\cite{hutchins1986direct}).

\vspace{0.65mm}
\bpstart{Semantics.}
Successfully retargeting a visualization must also consider data semantics.
Such decisions involve how new data fields should be mapped to fields from the old dataset, along with appropriate choices for encoding channels, axis and legend scales, and data transformations.
For instance, \texttt{division} is mapped to the facet encoding (Fig.~\ref{fig:teaser}(E)) given that no other encodings represent discrete data; \texttt{date} is mapped to the x-axis (Fig.~\ref{fig:teaser}(C)) to illustrate unemployment changes over time; and \texttt{unemployment} is mapped to the color legend (Fig.~\ref{fig:teaser}(D)), which would be inappropriate for the other fields.

After mapping new data fields to encoding channels, the designer must also refine any scales or transformations based on semantic overlap.
For example, the extent of the \texttt{mean\_unemployment} differs from that of the \texttt{mean\_temp} in the Seattle weather dataset, which must be reflected in the scale domain (Fig.~\ref{fig:teaser}(D)).
While the color scale does not need to be modified in this case, there are situations in which it might (\eg the new data field exhibits such a large variance that a logarithmic scale would be more appropriate).
In addition, the designer can reuse the aggregate transform for the \texttt{division} field (Fig.~\ref{fig:teaser}(B)), given that is represents discrete data.

Finally, the designer must also update axis labels / titles (Fig.~\ref{fig:teaser}(C)) and chart titles (Fig.~\ref{fig:teaser}(G)) to reflect the semantics and context of the new dataset.

\vspace{0.65mm}
\bpstart{Pragmatics.}
A retargeted visualization should also consider if the appropriate amount of information is conveyed for the target audience, as well as visual aesthetics.
For example, while a legend is not necessary for the original Altair ridgeline plot (Fig.~\ref{fig:teaser}(left)) given that it overlaps in context with the maximum daily temperature (x-axis), this assumption does not hold for the retargeted version (Fig.~\ref{fig:teaser}(right)), which should include a legend (Fig.~\ref{fig:teaser}(D)) to help the reader understand what color hue represents.

Other pragmatic decisions may include whether axis units should be included, if enough spacing is provided to prevent mark occlusion and ensure legibility, where text and legends should be positioned, how axis labels should be formatted, or how the chart should be modified for different modalities (\eg responsive visualization for smaller screens~\cite{hoffswell2020techniques}).
For charts with communicative aids such as annotations, decisions must be made as to whether they should be removed or not.
In this work, we primarily focus on supporting the LLM's analysis and refinement of syntax and semantics, given that pragmatic decisions are more difficult to automate and significantly depend on variable external factors (\eg audience, communication / analysis goals, context of viewing, visual aesthetic choices, \etc).

\section{Evaluating LLM-Assisted Retargeting}

\subsection{Methods}
We evaluate two strategies of LLM assistance.
The first is a baseline approach that queries an LLM with minimal context consisting of only the chart specification and summary of the new dataset.
The second is an LLM program synthesis pipeline that guides programmatic code generation with salient chart information (\eg visual encodings, types of transformations, \etc) to enhance reasoning and provide useful context, mitigating existing design challenges (\S\ref{sec:challenges}).
This pipeline consists of three sequential stages: syntax parsing, data mapping, and specification refinement.

\tableerroranalysis

\vspace{0.65mm}
\bpstart{Syntax Parsing.}
We parse the chart specification into its Abstract Syntax Tree (AST), enabling automated detection and analysis of dependencies that occur when deriving or transforming new variable names~\cite{xie2024waitgpt, gu2024blade}.
We then query the LLM to identify semantic roles of different variables and code sections, such as encoding channels (\eg axes, legends), data fields, and transformations.

\vspace{0.65mm}
\bpstart{Data Mapping.}
We provide the LLM with both a summary of the new dataset (\eg statistics, typing) and semantic roles from the syntax parsing stage to help it determine how the new dataset should be mapped to existing encodings (\eg based on typing and semantic similarity) and appropriately transformed.

\vspace{0.65mm}
\bpstart{Specification Refinement.}
Finally, we query the LLM with the new data mappings and code dependencies from the syntax parsing stage to assist with code generation and adaptation: removing stale transformations and encoding channels, as well as updating titles, axis labels (\eg units, formatting), encoding scales, and basic aesthetics (\eg filtering or aggregating data to prevent overplotting).

\vspace{1mm}
We evaluated both the LLM baseline and pipeline across 4 datasets~\cite{amzndataset, blsdataset, braindataset, imdbdataset} and 8 Python charts~\cite{mplboxplot, mplbarchart, mpldivergingbars, plotly, seabornheatmap, mplareachart, altair, seabornscatter} of varying complexity (different charting tools, chart types and encoding decisions, communication aids such as annotations, and data transformations), resulting in 64 retargeted visualizations (2 conditions $\times$ 4 datasets $\times$ 8 charts).
We used gpt-4o as the LLM.

\subsection{Results}
To evaluate the performance of retargeting output, we open-coded errors across each of the 64 charts, noting both syntactic issues with the specification as well as semantic and aesthetic errors in the visual output.
\textcolor{black}{Specifically, the first author began by manually inspecting each retargeted specification and its rendered chart to extract fine-grained errors.
Throughout this process, syntactic, semantic, and pragmatic axes (\S\ref{sec:challenges}) were used as thematic aids to help consider different dimensions of errors for completeness.
For example, the first author analyzed runtime, compilation, and linting errors to extract syntactic failures. 
They then analyzed visual components of the rendered chart (marks, axes, scales \& encodings, legends \& titles, \etc) to identify ambiguous or inappropriate semantics and reflect on the pragmatics of the design (legibility, formatting, positioning, unnecessary or missing chart information, \etc).
The first author coded the extracted error data to surface common types of errors (Table~\ref{table:errors}).
Initial codes were refined and merged via pilot-testing with a subset of the data to ensure reliability.}
In addition, we noted the severity of an error type if it prevented the retargeted chart from rendering or it was not possible to reasonably interpret what was being displayed (\eg significant overplotting due to lack of necessary transformations such as filtering or aggregation).

\figureheatmap

The results of our analysis our provided in Table~\ref{table:errors}.
\textcolor{black}{Out of the 32 chart instances per condition, 25 rendered in the baseline approach, but only 16 rendered with the LLM pipeline.}
Charts that failed to render were the result of syntactic errors, making it difficult or impossible to identify many of the downstream semantic or pragmatic decisions (\eg Fig.~\ref{fig:examples}(F)).
As a result, we normalized all semantic and pragmatic errors by the total number of rendered charts.
We found that the number of semantic and pragmatic errors for the pipeline were comparable to that of the baseline.
Both approaches were unable to adequately transform data or remove certain encodings altogether to prevent overplotting (\eg Fig.~\ref{fig:examples}(B,C,E)).
Further, the pipeline generated noticeably more syntactic errors, likely due to inference failures in the syntax parsing stage.
If the LLM fails to appropriately infer semantic roles, this information may be incorrect or incomplete, complicating subsequent queries in the pipeline.

To determine relationships between retargeting performance and the characteristics of the datasets and charts, we plotted a heatmap with the counts of errors (Fig.~\ref{fig:heatmap}).
We excluded counts from the LLM pipeline given its inability to render a noticeably larger number of charts compared to the baseline. 
We found that the number of errors increase for more complex charts, such as annotations or multiple layers (\eg regression lines as in Fig.~\ref{fig:examples}(D)).
In addition, both simple and complex datasets struggle with retargeting.
For instance, the IMDb movies dataset~\cite{imdbdataset} contains multiple columns that may require custom transformations prior to successful encoding.
As shown in Fig.~\ref{fig:examples}(E), although the movie \texttt{genre} is an appropriate fit to replace \texttt{species} in the original plot (Fig.~\ref{fig:examples}(D)), this mapping requires necessary transforms to unpack the arrays.
Compared with other datasets, the US unemployment data~\cite{blsdataset} contained the fewest number of columns, each of distinct type.
\textcolor{black}{Datasets without sufficient complexity (\ie not enough data fields or types) can necessitate custom transformations to derive new fields or deleting complex code dependencies to reduce the number of encodings.}

\figureexamples

\subsection{Design Recommendations}
Based on the performance of LLM-assisted retargeting, as well as the type and severity of different errors, we discuss our recommendations for future retargeting systems.

\vspace{0.65mm}
\bpstart{DR1: Enable Mixed-Initiative Assistance at Inflection Points.}
Fig.~\ref{fig:heatmap} reveals that at least one error occurred with each instance of retargeting.
An obvious result is that mixed-initiative interactions and interfaces will be necessary for some time to steer the LLM through the retargeting process (as with chart deconstruction tools~\cite{masson2023chartdetective}).
However, the inflection points for user interaction and refinement will be different according to the author's retargeting goals, dataset and specification complexity, and types of errors that may be encountered.
We expect the syntactic, semantic, and pragmatic axes for retargeting challenges (\S\ref{sec:challenges}) to serve as useful stages for users to interact with intermediate LLM output, mitigating cascading errors and restricting the type of code that is generated.

\vspace{0.65mm}
\bpstart{DR2: Surface Data Dependencies.}
Despite the LLM pipeline's relatively weak performance, surfacing dependencies between data fields, transformations, and encodings from the syntax parsing phase could allow designers to inspect and refine inference errors.
These corrections may help reduce subsequent semantic and pragmatic errors such as missing transformations or failing to update scale domains that use hard-coded values (Fig.~\ref{fig:teaser}(D)).
Further, operating on AST representations may help isolate portions of the specification that should not be changed without explicit approval (\eg specific transforms or encodings that are still desired), counteracting hallucinations or incorrect assumptions about the data.

\vspace{0.65mm}
\bpstart{DR3: Integrate Transformation Support.}
Prior work has explored LLM-assisted transformations with data tables to rapidly construct charts~\cite{wang2023data, 10.1145/3706598.3713296}, which generate precise scripts to manipulate and derive data columns.
However, transformation decisions in visualization retargeting are often less flexible since they are constrained by possible mappings between data fields from the old and new datasets.
As a result, future retargeting systems may benefit more from user-guided heuristics or flags that ease transformation inference.
For example, despite being prompted to consider filtering and aggregation transformations, overplotting persisted with both the LLM baseline (Fig.~\ref{fig:examples}(B)) and pipeline (Fig.~\ref{fig:examples}(C)).
Rather than generating a custom, one-off transformation script, the designer might be able to set explicit, reconfigurable flags on the retargeted output that coerces alignment with specific design constraints (\eg filter / aggregate the data, remove an encoding, facet, \etc).

\section{Limitations \& Future Work}

\textcolor{black}{Image-based retargeting approaches (\eg using vision models to analyze and manipulate SVG or bitmap images) might detect and correct visually salient errors (\eg overlapping marks) with greater accuracy.
Curated corpora~\cite{chenvisanatomy2024} can provide high-quality images from a large set of diverse charting tools for evaluation.
However, retargeting code provides other benefits: access to data transforms, portability of interactions and accessibility aids, and transparent retargeting edits via code diffs.
Rather, we expect that using code and image output together can leverage the benefits of both.
This approach can augment low-resource coding languages or new charting libraries that LLMs have not been adequately trained on.
Notational design~\cite{10.1007/3-540-44617-6_31} (\eg viscosity, hidden dependencies, abstraction) will likely affect performance as well (\eg more complex viscous languages that require many changes to achieve one retargeting goal).}

Newer reasoning models have been released since the start of this work (\eg Claude 3.7 Sonnet~\cite{claude}).
Although these models may reason more effectively about data dependencies and transformations, we expect that our analysis of retargeting errors can help characterize performance differentials between models and support future benchmarking.
\textcolor{black}{Syntactic and semantic errors may occur less frequently as models mature in their coding and reasoning abilities (\eg chain-of-thought~\cite{wei2022chain}). 
However, pragmatic mistakes are likely to persist given their reliance on user input and shifting analysis contexts (\eg tasks, goals, \etc). 
For instance, certain encodings or annotations, unit abbreviations / axis labels, and titles may or may not be necessary depending on the user's background and downstream viewing context (\eg analysis vs. communication).}


Our evaluation and analysis underscore the need for new interfaces and interaction paradigms to aid visualization retargeting.
These systems must balance the need between ensuring correct retargeting behavior and involving the designer where and when it is appropriate.
In particular, designers may lack familiarity with specific charting tools that could make it difficult for them to recognize syntactic or semantic errors in the specification, or know how to correct errors they have identified.
One possible solution might be to sequence retargeting operations deliberately, helping users isolate specific changes and relate them with the image output.

\acknowledgments{
This work was mostly performed during
an internship at Microsoft.
We thank the VIDA team at Microsoft Research for their thoughtful feedback and support.
}

\bibliographystyle{abbrv-doi}

\bibliography{bibliography}
\end{document}